\documentclass[10pt, conference, letterpaper]{IEEEtran}
\usepackage{algorithmicx}
\usepackage[ruled,vlined,linesnumbered]{algorithm2e}
\usepackage{hhline}
\usepackage{amsmath,mathtools}
\usepackage{amsfonts,amssymb}
\usepackage{mathrsfs}
\usepackage{gensymb} 
\usepackage{graphicx} 
\usepackage{multirow}
\usepackage{enumitem,color}
\usepackage{algpseudocode}
\usepackage{makecell}
\usepackage{booktabs}

\usepackage{titlesec}
\titlespacing*{\section}{1pt}{0.5ex}{0.5ex}
\titlespacing*{\subsection}{1pt}{0.5ex}{0.5ex}
\titlespacing*{\subsubsection}{1pt}{0.5ex}{0.5ex}
\titleformat{\subparagraph}[runin]{\normalfont\normalsize\bfseries}{\thesubparagraph}{1em}{}

\begin{document}
%
\title{\huge \emph{AdaSlicing}: Adaptive Online Network Slicing under Continual Network Dynamics in Open Radio Access Networks \vspace{-0.1in}}

\author{\IEEEauthorblockN{Ming Zhao, Yuru Zhang, Qiang Liu \vspace{-0.16in}}\\
\IEEEauthorblockA{
University of Nebraska-Lincoln\\
qiang.liu@unl.edu
}\vspace{-0.4in}
\and
\IEEEauthorblockN{Ahan Kak, Nakjung Choi \vspace{-0.16in}}\\
\IEEEauthorblockA{
Nokia Bell Labs\\
nakjung.choi@nokia-bell-labs.com
}\vspace{-0.4in}
}

\maketitle

\begin{abstract}
Open radio access networks (e.g., O-RAN) facilitate fine-grained control (e.g., near-RT RIC) in next-generation networks, necessitating advanced AI/ML techniques in handling online resource orchestration in real-time.
However, existing approaches can hardly adapt to time-evolving network dynamics in network slicing, leading to significant online performance degradation.
In this paper, we propose \emph{AdaSlicing}, a new adaptive network slicing system, to online learn to orchestrate virtual resources while efficiently adapting to continual network dynamics.
The AdaSlicing system includes a new soft-isolated RAN virtualization framework and a novel AdaOrch algorithm.
We design the AdaOrch algorithm by integrating AI/ML techniques (i.e., Bayesian learning agents) and optimization methods (i.e., the ADMM coordinator).
We design the soft-isolated RAN virtualization to improve the virtual resource utilization of slices while assuring the isolation among virtual resources at runtime.
We implement AdaSlicing on an O-RAN compliant network testbed by using OpenAirInterface RAN, Open5GS Core, and FlexRIC near-RT RIC, with Ettus USRP B210 SDR.
With extensive network experiments, we demonstrate that AdaSlicing substantially outperforms state-of-the-art works with 64.2\% cost reduction and 45.5\% normalized performance improvement, which verifies its high adaptability, scalability, and assurance.
\end{abstract}


\section{Introduction}
\label{sec:introduction}

Network slicing is a key technique in 5G and Beyond to cost-efficiently and flexibly support emerging mobile applications, e.g., autonomous driving~\cite{bagheri20215g}, extended reality~\cite{guan2022deepmix}, and Internet of Things~\cite{qiu2020edge}.
With the advanced technology of infrastructure virtualization, multiple virtual networks (i.e., slices) can be concurrently instantiated and operated on common physical infrastructures (e.g., base stations and switches), with assured resource and performance isolation.
By tailoring the parameters and resources of each slice, mobile network operators (MNOs) can effectively meet the diversified performance needs of slice tenants, such as end-to-end latency, security, and reliability.
In network slicing, resource orchestration serves as the key role to dynamically manage virtualized resources~\cite{liu2021onslicing, liu2022atlas} in multiple technical domains (e.g., radio spectrum) to all slices for assuring their service-level agreements (SLAs). 

Existing in-use orchestration solutions~\cite{marquez2018should, shi2021adapting,salvat2018overbooking} heavily rely on human expertise throughout the life-cycle of each slice (e.g., performance modeling and fine-tuning), where orchestration actions are optimized in the coarse granularity, such as every hour.
With the increasing momentum of open network initiatives (e.g., O-RAN~\cite{o-ran,polese2023understanding}) and an emphasis on a unified software approach to simplify operations amidst increasing complexity (e.g., UNEXT~\cite{unext}), next-generation networks will expose high-dimensional states (e.g., thousands if not more) and allow nearly real-time control (e.g., subseconds), which enables more fine-grained orchestration in network slicing for further exploiting resource multiplexing~\cite{marquez2018should}.
To tackle the complex fine-grained orchestration problem, machine learning (ML) techniques~\cite{lecun2015deep} have been increasingly explored, such as deep reinforcement learning~\cite{lillicrap2015continuous} and Bayesian learning~\cite{liu2021onslicing, liu2022atlas}, and achieved great improvements, in terms of performance, autonomy, and scalability.

However, we found that existing works can hardly adapt to time-varying dynamics in network slicing, which can result in substantial performance degradation (e.g., violated SLAs and soared resource usage), especially during online resource orchestration.
Generally, existing works rely on deep neural network (DNN)-parameterized agents to manage resource orchestration for all slices, where their fixed input and output space\footnote{Although there are several DNN architectures with flexible inputs (e.g., LSTM and GNN), their training complexity and sample efficiency are widely concerned to be used for online network management~\cite{liu2021onslicing, mao2019learning}.} limit the adaptability to diverse network dynamics.
On the one hand, the active slices in the network are not stationary. 
Independent slice tenants may operate their slices dynamically, such as starting and stopping slices at different times throughout the day. 
On the other hand, the traffic pattern and application characteristics of each slice may change and evolve over time.
As a result, existing DNN-parameterized agents have to be retrained and refined, leading to delayed adaptation to time-varying network dynamics.

In this paper, we propose a new adaptive network slicing system (\emph{AdaSlicing}), to online learn while efficiently adapting to time-varying network dynamics.
The key idea is to integrate AI/ML techniques and optimization methods during online resource orchestration.
We design a new AdaOrch algorithm to minimize the total operating cost of supporting all slices, while assuring the performance requirements defined by their SLAs.
On the one hand, we design a Bayesian learning agent to handle the resource orchestration for each slice, which will be continually updated with accumulated online experiences.
On the other hand, we design a coordinator to coordinate all active slices in regard to the infrastructure capacity of virtual resources at runtime.
Moreover, we design a new soft-isolated RAN virtualization to improve the virtual resource utilization of slices while assuring the isolation among virtual resources at runtime.
We implemented AdaSlicing on an end-to-end O-RAN compliant network testbed by using OpenAirInterface RAN, Open5GS Core, and FlexRIC near-RT RIC, with Ettus USRP B210 SDR.
With the extensive network experiments, we demonstrate that AdaSlicing can reduce 64.2\% total operating cost while improving 45.5\% normalized performance of slices, as compared to state-of-the-art solutions.

Overall, we propose \emph{AdaSlicing}, a new adaptive online network slicing system, that can flexibly adapt to diverse time-varying network dynamics. The detailed contributions are:
\begin{itemize}[leftmargin=*]
    \item We design a new soft-isolated RAN virtualization framework that substantially improves virtual resource utilization of slices at runtime. 
    \item We design a new AdaOrch algorithm that online learns and orchestrates virtual resources for slices with assured SLAs.
    \item We implement the AdaSlicing system on an O-RAN compliant mobile network testbed, with multiple slices and users.
    \item We conducted extensive experiments to evaluate AdaSlicing, in terms of adaptability and scalability.
\end{itemize}

\section{$AdaSlicing$ Overview}
\label{sec:adaslcing}

\begin{figure}[!t]
	\centering
	\includegraphics[width=3.3in]{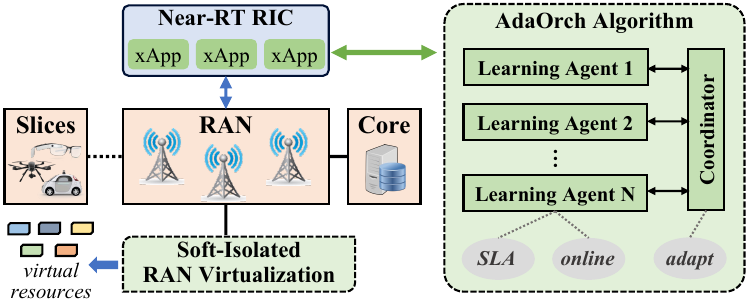}
	\caption{\small The overview of AdaSlicing.} 
	\label{fig:overview}
\end{figure}
In Fig.~\ref{fig:overview}, we overview the AdaSlicing system under the architecture of open radio access networks.
It includes multiple slices, radio access network (RAN), and core network (CN), where the near-RT RIC hosts a wide range of xApps, such as performance monitoring xApps and the AdaOrch algorithm xApp.
In AdaSlicing, two primary components are the soft-isolated RAN virtualization and the AdaOrch algorithm.

We design the \emph{soft-isolated RAN virtualization} to improve the utilization of virtual resources at runtime while assuring the isolation among virtual resources (See Sec.~\ref{sec:ran_virtualization}).
Different from existing hard-isolated RAN virtualization, our key idea is to enable the sharing of unused virtual resources among slices, before the virtual-to-physical mapping, at runtime.
We create two new kinds of virtual resources, including the soft-isolated virtual resource block (svRB) and sharing weight (SW), which can be dynamically orchestrated to all slices by the AdaOrch algorithm.
Note that all unused virtual resources are shared proportionally according to the SW value of slices, which improves resource utilization at runtime and creates performance interdependence among all slices.
In other words, in addition to the svRB, the shared virtual resources of a slice from the sparse vRB pool depend on not only its SW value but also the SW values of all other active slices. 

We design the \emph{AdaOrch algorithm} to online orchestrate virtual resources for all slices under time-varying network dynamics (See Sec.~\ref{sec:solution}).
Different from existing orchestration solutions, our key idea is to integrate AI/ML techniques and optimization methods to improve adaptability while maintaining the autonomy of the AdaOrch algorithm.
Specifically, it includes multiple learning agents, where each agent corresponds to online learning and orchestrates virtual resources for a slice.
Note that, the AdaOrch algorithm can support heterogeneous learning agents, in terms of adopted AI/ML techniques (e.g., Bayesian learning and multi-armed bandit), as long as they match the same input/output space and also follow necessary training processes.
It also includes a coordinator, designed based on the alternating direction method of multipliers (ADMM), to coordinate the infrastructure capacity of virtual resources for all slices.
During online orchestration, each learning agent observes its local network context and makes the orchestration action for its corresponding slice to meet the performance requirement defined by the slice SLA.
At runtime, only active slices are involved and iteratively communicated with the coordinator to achieve a consensus of resource capacity.
On the one hand, these learning agents will be online updated according to newly obtained online experiences, which assures the continual learning capability to adapt to potential intra-slice network dynamics.
On the other hand, the coordinator can support an arbitrary number of learning agents during each orchestration slot, which can flexibly adapt to possible inter-slice network dynamics.

\section{System Model}
\label{sec:problem}
We consider an O-RAN compliant mobile network, including the core network (CN) and radio access network (RAN) with multiple base stations (BSs)\footnote{Without loss of generality, we focus on the resource orchestration problem in a single base station, where the AdaOrch algorithm can be easily extended to support the scenario of multiple base stations.} and network slices. 
Slice tenants establish the service level agreement (SLA) with the mobile network operator (MNO) to support their slice users, with predefined performance requirements, such as the maximum latency and minimum throughput.
The MNO creates multiple xApps in the near-RT RIC to dynamically orchestrate the virtual resources for all slices in the fine time granularity (e.g., every second). 
Here, we focus on two kinds of virtual resources: 1) the soft-isolated virtual resource blocks (svRB), and 2) the sharing weight (SW), which are detailed in Sec.~\ref{sec:ran_virtualization}.
We denote $\mathcal{I}$ as the set of network slices, where $i \in \mathcal{I}$ denotes the $i$-th slice. 

\textbf{Action Space.} 
Denote $x_i^{(t)}$ and $w_i^{(t)}$ as the number of svRBs and the SW value of the $i$-th slice at the $t$th time slot, respectively. 
For the sake of simplicity, we further denote $\mathcal{X}^{(t)} = \{x_i^{(t)}, \forall i\in \mathcal{I}\}$ and $\mathcal{W}^{(t)} = \{w_i^{(t)}, \forall i\in \mathcal{I}\}$ as the set of these orchestration actions, respectively.
Therefore, we define the action space of resource orchestration to the $i$-th slice as
\begin{equation}
    A_i^{(t)} = \{x_i^{(t)},w_i^{(t)}\}.
\end{equation}

\textbf{Performance Model.}
Under the soft-isolated RAN virtualization (see Sec.~\ref{sec:ran_virtualization}), the final experienced vRBs of a slice depend on not only its orchestration action but also that of other slices (particularly their SW values).
Hence, we define the performance function of the $i$-th slice as
\begin{equation}
    P_i^{(t)} = f(x_i^{(t)},w_i^{(t)}|s_i^{(t)}),
\end{equation}
where we introduce $s_i^{(t)}=\sum\nolimits_{\substack{j \in \mathcal{I}, j \neq i}} w_j^{(t)}$ as the aggregated SW value of all other active slices.
Here, the performance metrics of each slice can be multi-dimensional and heterogeneous, such as end-to-end latency and reliability.
Here, more network context might be incorporated, if relevant and needed, to better represent the performance function of slices.

\textbf{Cost Model.}
From the MNO perspective, the operating cost of running a slice highly depends on its usage of virtual resources.
Here, we define operating cost of the $i$-th slice as
\begin{equation}
 U_i^{(t)} = {U_H} \cdot {x_{i}^{(t)}}+{U_S} \cdot {w_i^{(t)}},
\end{equation}
where $U_H$ and $U_S$ denote the unit cost of svRB and SW, respectively.

\textbf{Problem.}
The objective is to minimize the total cost of supporting all slices, while meeting their performance requirements defined by slice SLAs.
Therefore, we formulate the resource orchestration problem in network slicing as
\begin{align}
 \mathcal{P}_0:&\quad \min_{\{ \mathcal{X}^{(t)},\mathcal{W}^{(t)} \}}\quad \sum\nolimits_{i\in\mathcal{I}} U_i^{(t)}\\
 s.t. \quad &C1: \quad f(x_i^{(t)},w_i^{(t)}|s_i^{(t)}) \ge Q_{i},{\forall}i \in \mathcal{I},\\
 &C2: \quad 0 \le w_i^{(t)}\le 1,{\forall}i \in \mathcal{I}, \\
 &C3: \quad 0 \le x_i^{(t)}\le H, {\forall}i \in \mathcal{I},\\
 &C4: \quad 0 \le \sum\nolimits_{i\in\mathcal{I}}x_i^{(t)} \le H.
 \label{p0}
\end{align}
Here, the constraint $C1$ assures that the performance of each slice can be maintained, where the performance threshold of the $i$-th slice is denoted as $Q_i$.
The constraint $C2$ and $C3$ define the boundary of the action space, including the maximum number of svRBs.
Moreover, the constraint $C4$ assures that the infrastructure capacity (denoted by $H$) will not be exceeded at runtime.

\textbf{Challenge.} The challenge of addressing the above problem is mainly two-fold.
On the one hand, the multi-dim performance of slices relates to a wide range of factors (e.g., traffic pattern and channel quality) and can hardly be represented in closed-form with respect to resource orchestration actions. 
Moreover, the performance function of slices could change over time, depending on their time-evolving application characteristics.
As a result, it is difficult to use conventional math modeling approaches to represent the complex and time-evolving performance function.
On the other hand, the active slices are non-stationary, depending on the operation strategy of independent slice tenants, such as starting and stopping slices at different times throughout the day. 
As a result, it is inefficient to handle the dynamics of active slices by using only parameterized agents with fixed input and output spaces.

\section{Proposed Solution}
\label{sec:solution}
In this section, we develop the AdaOrch algorithm to efficiently solve the resource orchestration problem.
The fundamental idea is to integrate AI/ML techniques and optimization methods to address the aforementioned adaptability challenge.
On the one hand, we design a learning agent (based on the Bayesian optimization framework) that focuses on the resource orchestration of each slice, which will be online updated with accumulated experiences to track its potentially time-evolving performance function.
On the other hand, we design a coordinator (based on the alternating direction method of multipliers (ADMM) framework) to coordinate all active slices in regard to the infrastructure capacity of virtual resources.
In addition, we design the interface (e.g., state and action space and training mechanism) to enable the convergent interaction among learning agents and the coordinator towards the objective of adaptive network slicing.

Specifically, we first decouple the original problem $\mathcal{P}_0$ into multiple subproblems and a coordination problem, by leveraging the ADMM framework.
We design each subproblem to correspond to the resource orchestration of individual slices, and the coordination problem to coordinate and assure the infrastructure capacity of virtual resources (i.e., $C4$).
The coordination problem turns out to be convex and can be efficiently solved by off-the-shelf optimization toolboxs.
To solve individual subproblems, we design a constrained Bayesian optimization method to online learn and orchestrate virtual resources for individual slices, while assuring their performance requirements.
During online orchestration, these subproblems and the coordination problem will be solved alternatively, and eventually achieve a convergent optima, which acts as the final resource orchestration for all slices.

\subsection{Problem Decomposition}
First, we introduce auxiliary variables $z_i^{(t)}$ and enforce additional constraints by letting $z_i^{(t)} = x_{i}^{(t)},{\forall i \in \mathcal{I}}$. 
Then, we can reformulate the problem $\mathcal{P}_0$ into the following problem 
\begin{align}
    \mathcal{P}_1:&\quad  \min_{\{ \mathcal{X}^{(t)},\mathcal{Z}^{(t)},\mathcal{W}^{(t)} \}}\quad \sum\nolimits_{i\in \mathcal{I}} U_i^{(t)}\\
    s.t.  \quad & C1,\; C2,\; C3,\\
                &C4: \quad 0 \le \sum\nolimits_{i\in\mathcal{I}}z_i^{(t)} \le H,\\
                &C5: \quad  x_{i}^{(t)} = z_i^{(t)},{\forall i \in \mathcal{I}},
    \label{p1}
\end{align}
where we rewrite the constraint $C4$ in $\mathcal{P}_0$ (which relates to $\mathcal{X}^{(t)}$) into a new constraint in this problem $\mathcal{P}_1$, which relates to only $\mathcal{Z}^{(t)}$.
Here, we denote $\mathcal{Z}^{(t)} = \{z_i^{(t)} , \forall i \in \mathcal{I}\}$ as the set of auxiliary variables.
The above problem reformulation aims to decouple the connection between the optimization variables of $\mathcal{X}^{(t)}$ in the original constraint $C4$ in $\mathcal{P}_0$, which will facilitate the separation of individual $x_{i}^{(t)}$ in each slice.
Note that, the optimization variable $w_i^{(t)}$ in each slice is not included by any constraints, except its boundary between 0 and 1.
In the reformulated problem $\mathcal{P}_1$, we have three kinds of optimization variables, i.e., $\mathcal{X}^{(t)}$, $\mathcal{Z}^{(t)}$, and $\mathcal{W}^{(t)}$.

Second, we derive the augmented Lagrangian function of $\mathcal{P}_1$ with scaled dual variables as
\begin{equation}
\mathcal{L}(\mathcal{X}, \mathcal{W}, \mathcal{Z}, \mathcal{Y}) = \sum\limits_{i\in \mathcal{I}}(U_i^{(t)}+\frac{\rho}{2}\left \| x_i^{(t)}-z_i^{(t)}+y_i^{(t)} \right\|^2),
\label{Lagrangian}
\end{equation}
where $\rho$ is a positive constant, and $\mathcal{Y}^{(t)} = \{y_i^{(t)} , \forall i \in \mathcal{I}\}$ is the set of scaled dual variables. 
Based on the ADMM framework~\cite{boyd2011distributed}, we can solve the problem $\mathcal{P}_1$ by alternatively addressing the following problems:
\begin{align}
    \mathcal{P}_2: \quad &  \mathcal{X}^{(t+1)}, \mathcal{W}^{(t+1)} = \nonumber \\ 
    & \arg\min_{\mathcal{X}^{(t)}, \mathcal{W}^{(t)}\in \{C_1,C_2,C_3\}}  \mathcal{L}(\mathcal{X}, \mathcal{W}, \mathcal{Z}^{(t)}, \mathcal{Y}^{(t)}), \\
    \mathcal{P}_3: \quad &  \mathcal{Z}^{(t+1)} = \nonumber \\ 
    & \arg\min_{\mathcal{Z}^{(t)} \in C_4}  \mathcal{L}(\mathcal{X}^{(t+1)}, \mathcal{W}^{(t+1)}, \mathcal{Z}, \mathcal{Y}^{(t)}),
\end{align}
and updating the dual variables as follows:
\begin{align}
    y_i^{(t+1)} = y_i^{(t)} + (x_i^{(t+1)}-z_i^{(t+1)}), {\forall}i \in \mathcal{I}.
    \label{update_y}
\end{align}
Here, problem $\mathcal{P}_2$ involves the actual resource orchestration for all slices, under the given auxiliary and dual variables in the last iteration.
Then, problem $\mathcal{P}_3$ relates to the update of auxiliary variables, depending on the optimized orchestration actions of problem $\mathcal{P}_2$.
Next, dual variables can be updated with the optimized orchestration actions and auxiliary variables.
This iteration continues until the convergence of variables.

\subsection{The Design of Coordinator}
In the context of the AdaSlicing system, we centralize the solving of problem $\mathcal{P}_3$ and the update of dual variables into the coordinator.
Specifically, we rewrite the problem $\mathcal{P}_3$ as
\begin{align}
    \mathcal{P}_4:&\quad  \min_{z_{i}^{(t)}}\quad \sum\nolimits_{i\in\mathcal{I}}\left \| x_i^{(t)}-z_i^{(t)}+y_i^{(t)} \right\|^2\\
    s.t. \quad & \quad  0 \le \sum\nolimits_{i\in\mathcal{I}}z_i^{(t)} \le H,
    \label{p4}
\end{align}
where the first part of augmented Lagrangian is irrelevant to this optimization of auxiliary variables and thus omitted.
We observe that this problem $\mathcal{P}_4$ is a standard quadratic integer programming problem, which is convex. 
Hence, we can utilize the off-the-shelf optimization toolbox to efficient solve it, such as CVX~\cite{diamond2016cvxpy,agrawal2018rewriting}. 
By solving problem $\mathcal{P}_4$, we obtain the updated auxiliary variables $\mathcal{Z}^{(t+1)}$, and then update the dual variables accordingly.

\subsection{The Design of Learning Agents}
In the context of the AdaSlicing system, we solve the problem $\mathcal{P}_2$ in these learning agents.
This is based on the observation that, the problem $\mathcal{P}_2$ is fully separable with respect to each slice, where the optimization of $x_{i}^{(t)}$ can be fully conducted in the $i$-th slice without any connections with other slices.
For the sake of simplicity, we re-express the problem $\mathcal{P}_2$ into the following child problem $\mathcal{P}_5$ in the $i$th slice
\begin{align}
    \mathcal{P}_5:&\quad  \min_{\{x_{i}^{(t)},w_{i}^{(t)}\}}\quad U_i^{(t)}+\frac{\rho}{2}\left \| x_i^{(t)}-z_i^{(t)}+y_i^{(t)} \right\|^2\\
    s.t. \quad &\bar{C1}: \quad  f(x_i^{(t)},w_i^{(t)}|s_i^{(t)}) \ge Q_i,\\
    &\bar{C2}: \quad  0 \le w_i^{(t)}\le 1, \\
    &\bar{C3}: \quad  0 \le x_i^{(t)}\le H,
    \label{p5}
\end{align}
where all constraints are rewritten with respect to only the $i$th slice.
Due to the unknown and potential time-evolving performance function in constraint $\bar{C1}$, it is challenging to solve the problem with conventional optimization based methods, such as linear and convex optimization~\cite{boyd2010convex}.

\textbf{Constrained Bayesian Optimization.} 
Here, we design a new constrained Bayesian optimization method to address the above child problem $\mathcal{P}_5$.
Bayesian optimization~\cite{frazier2018tutorial,snoek2012practical} is the state-of-the-art global optimization framework, and is particularly efficient in handling blackbox problems with expensive querying costs.
It is an iterative searching process, including a probabilistic surrogate model and an acquisition function.
In each iteration, the surrogate model, e.g., Gaussian process~\cite{rasmussen2003gaussian}, is trained to approximate the uncertainty of the black-box function, e.g., the performance function of slices $f(x_i^{(t)},w_i^{(t)}|s_i^{(t)})$, according to previous experiences, e.g., orchestration-to-performance pairs.
Then, the acquisition function, e.g., expected improvement (EI)~\cite{brochu2010tutorial}, estimates the utility of different actions $\{x_{i}^{(t)},w_{i}^{(t)}\}$ while balancing the exploration and exploitation.
The next action can be determined by maximizing the acquisition function, under the boundary of action space (e.g., $\bar{C2}, \bar{C3}$).
Along with the accumulation of online experiences, the surrogate model will be updated to be more accurate to represent the blackbox function, which guides the selection of future orchestration actions towards the optima.

\emph{\underline{Barrier Method.}}  
In the child problem $\mathcal{P}_5$, its objective function is deterministic with a closed-form expression, while its constraint $\bar{C1}$ is the unknown blackbox function.
As vanilla Bayesian optimization hardly handle complex constraints (i.e., $\bar{C1}$), we use the barrier method~\cite{boyd2010convex} to convert the child problem $\mathcal{P}_5$ into unconstrained one.
Specifically, we add a penalty function to the objective function, and rewrite the child problem $\mathcal{P}_5$ as
\begin{align}
    \mathcal{P}_6: \;\;\; \quad  \min_{\{x_{i}^{(t)},w_{i}^{(t)},s_{i}^{(t)}\}}\quad  & U_i^{(t)}+\frac{\rho}{2}\left \| x_i^{(t)}-z_i^{(t)}+y_i^{(t)} \right\|^2 +\phi_i \nonumber \\
     s.t.  & \quad \bar{C2}, \bar{C3},
    \label{p6}
\end{align}
where $\phi_i = -log(-(Q_i - f(x_i^{(t)},w_i^{(t)}|s_i^{(t)}))$ is the log barrier penalty.


\emph{\underline{Gaussian Process.}}
We adopt Gaussian Process (GP) as the surrogate model to approximate the time-evolving performance function of the slice, which is based on its superior advantages of sample efficiency and robustness.
Gaussian Process \cite{williams1995gaussian} is a probabilistic, sample-efficient and wide-adopted non-parametric machine learning technique. 
With slight abuse of notation, we denote $\textbf{x} = [x_i^{(t)},w_i^{(t)},s_i^{(t)}]$ as the combination of the orchestration action of this slice and aggregated SW of other slices.
Gaussian Process can be defined by its mean function $ m(\textbf{x}) $ and covariance function $k(\textbf{x},\textbf{x}') $ as:
\begin{equation}
m(\textbf{x}) = \mathbb{E}[f(\textbf{x})],
\end{equation}
\begin{equation}
k(\textbf{x},\textbf{x}') = \mathbb{E}[(f(\textbf{x}) - m(\textbf{x}))(f(\textbf{x}') - m(\textbf{x}'))],
\end{equation}
where $ f(\textbf{x}) $ represents the unknown performance function of the slice to learn (i.e., $f(x_i^{(t)},w_i^{(t)}|s_i^{(t)})$). 
Here, $ k $ is the kernel function, which determines the smoothness and other properties of the functions drawn from the Gaussian Process.
For any new input $ \textbf{x}^* $, GP can generate in a normal distribution, with its mean denoted by $ \mu(\textbf{x}^*) $ and its variance by $ \sigma^2(\textbf{x}^* $), which will be utilized by the acquisition function when selecting the next action.

To track the potential time-evolving performance function of slices, we introduce a fixed-size reply buffer for the GP during the online orchestration.
Instead of using all observed experiences, we regress the GP by sampling experiences from the reply buffer, based on the priority of experiences.
When a new online experience is obtained, it will be queued into the reply buffer with the highest sampling priority.
Note that, we decay the priority of an experience according to its age-of-information, which will make the latest experiences to be sampled more frequently.
If the reply buffer is full, the oldest experience will be removed, which helps to track the time-evolving performance function of slices.
In the meantime, the fixed size of the reply buffer ensures that the computation complexity of GP regression is constant and can be performed in real-time.

\emph{\underline{Acquisition Function.}}
For the acquisition function, there is a wide range of candidates, such as lower confidence bound (LCB), expected improvement (EI), and probability of improvement (PI).
For example, EI aims to maximize the expected improvement under to-date observations, while balancing the exploration and exploitation.
However, existing works~\cite{vasconcelos2019no, hoffman2011portfolio} show that acquisition functions may fall into local optima under different blackbox functions, i.e., there is no one-fit-all acquisition function.
Hence, we adopt the $gp\_hedge$~\cite{hoffman2011portfolio} strategy in the AdaSlicing system. 
Its basic idea is to dynamically choose one of the candidate acquisition functions during the iterations of Bayesian optimization.
The selection of acquisition functions is optimized by using an online multi-armed bandit algorithm, which demonstrated promising performance compared to individual fixed acquisition functions.

\section{Soft-Isolated RAN Virtualization}
\label{sec:ran_virtualization}

\begin{figure}[!t]
	\centering
	\includegraphics[width=3in]{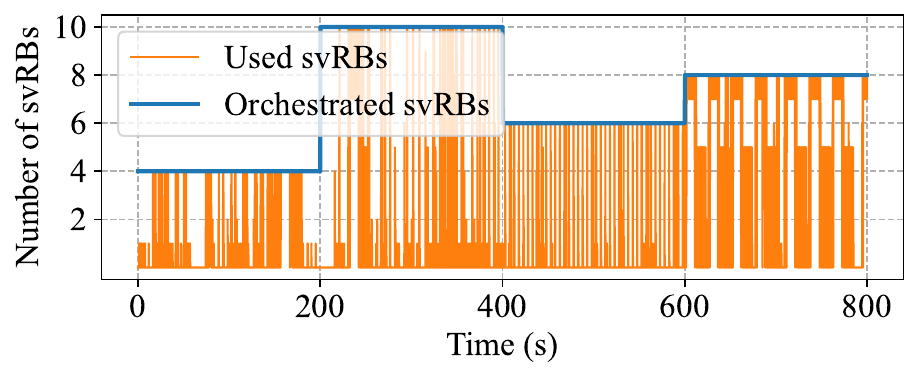}
	\vspace{-0.05in}\caption{\small An example of runtime utilization of vRBs under different applications. Here, we run mixed applications before 400s, only watch live video in [400s, 600s], and then perform speedtest.}
	\label{fig:total_vs_request_svrbs}
\end{figure}

In this section, we introduce the soft-isolated RAN virtualization in the AdaSlicing system.

RAN virtualization is the foundation of network slicing technique to virtualize physical infrastructures (e.g., base stations) into virtual radio resources (e.g., vRBs) and implement virtual resources at runtime.
During online resource orchestration, these virtual resources can be flexibly orchestrated to different network slices periodically (e.g., every minute or hour in conventional networks).
At runtime, virtual resources of slices will be mapped to physical resources (e.g., PRBs/RBGs) by using virtual-to-physical mapping~\cite{schmidt2021flexric, liu2019virtualedge,pham2023hexric}.
With the mapped physical resources, each slice will conduct its intra-slice user scheduling in every millisecond.
The objective of RAN virtualization is to achieve high isolation (e.g., resource) and low overhead (e.g., computation).
On the one hand, high isolation ensures that virtual resources are isolated with each other, and thus provides performance isolation among slices.
Here, we denote RAN virtualization as \emph{hard isolated} when virtual resources are only exclusively mapped to physical resources, e.g., a vRB always corresponds to a set of PRBs in the MAC layer.
On the other hand, low overhead ensures that the implementation of virtual resources can be performed at a minimal computation time.

\begin{figure}[!t] 
  \begin{minipage}[t]{0.24\textwidth}
    \centering
    \includegraphics[width=1.7in]{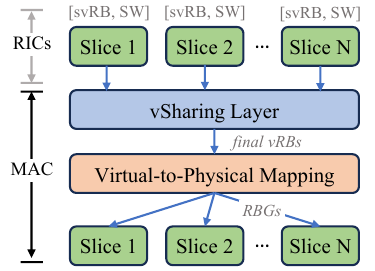}
    \caption{The architecture of soft-isolated RAN virtualization.}
    \label{fig:virtualization}
  \end{minipage}
  \begin{minipage}[t]{0.24\textwidth}
    \centering
    \includegraphics[width=1.7in]{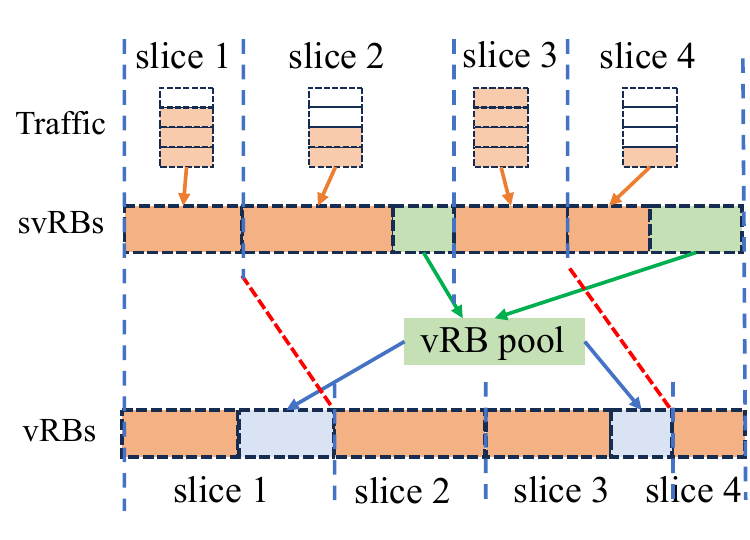}
    \caption{An example of soft-isolated RAN virtualization.}
    \label{fig:virtualization_example}
  \end{minipage}
\end{figure}

In the context of open radio access networks (e.g., O-RAN), the virtual resources of network slices can be orchestrated in fine time granularity (e.g., as low as subseconds via near-RT RIC) to better track their time-varying traffic variations.
However, we observe that the orchestrated virtual resources of slices are usually under-utilized in the scenario of hard-isolated RAN virtualization.
Fig.~\ref{fig:total_vs_request_svrbs} shows the number of vRB utilization of a network slice under different virtual resources and slice applications in a mobile network testbed.
It can be seen that, the orchestrated vRBs are not fully utilized due to different patterns of slice traffic, where only active speedtest may saturate the orchestrated vRBs.
This resource under-utilization can be attributed to 1) slices are usually orchestrated to have sufficient virtual resources to accommodate their peak traffic until the next orchestration slot; 2) their virtual resources are exclusively mapped and cannot be shared with other slices, even if they are not fully used at runtime.

Therefore, we propose a soft-isolated RAN virtualization, as shown in Fig.~\ref{fig:virtualization}, to improve the runtime utilization of virtual resources while assuring their isolation.
Soft-isolation also serves as an enabler for the 3GPP-defined network resource model~\cite{3gpp.28.541}, which is considered the \emph{de facto} standard for RAN slice resource management. 
The key idea is to enable the sharing of unused virtual resources among slices, before the virtual-to-physical mapping, at runtime.
Specifically, we design a new virtual resource sharing (vSharing) layer to share expectantly excessive virtual resources among slices, where the inputs are the orchestration action of all slices and the outputs are the number of vRBs of all slices.
First, we create two new kinds of virtual resources, including soft-isolated virtual resource blocks (svRBs) and sharing weights (SWs).
Here, we assume that a virtual resource block (vRB) always corresponds to a fixed set of physical resources, i.e., one downlink resource block group (RBG) and one uplink physical resource block (PRB), in the virtual-to-physical mapping.
Second, based on the orchestrated number of svRBs of each slice, we estimate the expectantly needed number of vRBs for intra-slice user scheduling, as illustrated in Fig.~\ref{fig:virtualization_example}.
This can be determined by aggregating all user traffic (e.g., as stored in the per-user RLC buffer) associated with a given slice.
Third, we determine the total number of unused vRBs of all slices and build a vRB pool, where we skip these slices whose expectantly needed vRBs exceed their orchestrated svRBs.
Fourth, we proportionally share the vRB pool to these slices with overflowed user traffic, according to their SWs.
Thus, we can obtain the final number of vRBs for all slices, including 1) for slices with overflowed traffic: summing up their orchestrated svRBs and the shared unused vRBs; 2) for slices without overflowed traffic: their expectantly needed vRBs.
Finally, the virtual-to-physical mapping will be invoked to map the final vRBs of slices to physical resources (i.e., PRBs and RBGs).

\section{System Implementation}
\label{sec:implementation}

\begin{figure}[!t]
	\centering
	\includegraphics[width=3in,height=1.7in]{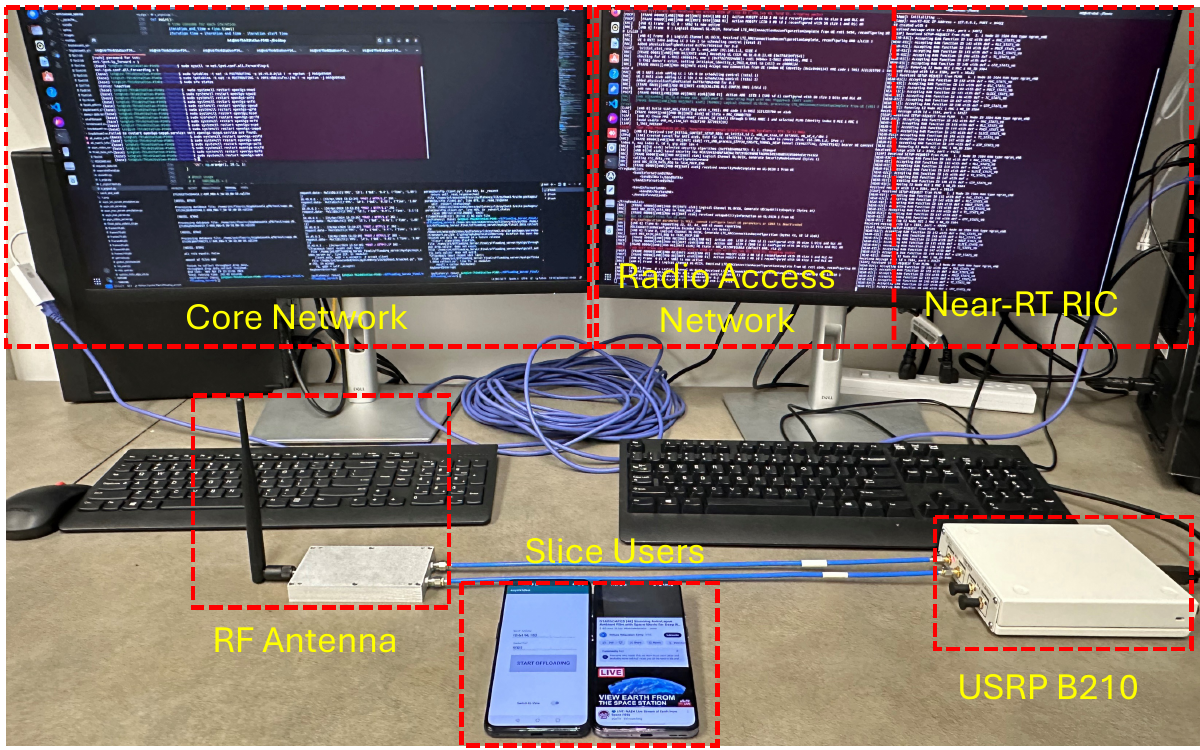}
	\caption{\small The overview of AdaSlicing testbed.}
	\label{fig:hardware}
\end{figure}
In this section, we describe the testbed implementation of the AdaSlicing system, including hardware, software, and architecture, and introduce the application of slices.

\subsection{Testbed Specifications}
We implement the AdaSlicing system on an end-to-end network slicing testbed, including the radio access network, core network, and near-real-time RIC, as illustrated in Fig.~\ref{fig:hardware}. 
We implement the RAN by using OpenAirInterface (OAI)~\cite{openairinterface5g} (v2022.41).
The RAN is hosted in an Intel i7-14700K desktop (64G RAM) with a low-latency kernel of Ubuntu 22.04, which connects an Ettus USRP B210 as the RF front-end.
We operate the base station at band 7 with 10MHz radio bandwidth (i.e., 50 physical resource blocks).
We use a Faraday cage to containerize all smartphones for eliminating other radio interference.
We implement the CN with Open5GS~\cite{open5GS} (v2.7.0), which is hosted in another Intel i7-10700 desktop (32G RAM).
The RAN and CN is connected with 1Gbps Ethernet cable.
We implement the near-RT RIC with FlexRIC~\cite{schmidt2021flexric} (v1.0.0), which supports 4G slicing capability.
We implement the AdaOrch algorithm with Python 3.11, which runs on the CN desktop.
We adopt the scikit-optimize (v0.10.1) with the Gaussian Process estimator, which relies on the \emph{GaussianProcessRegressor} module in \emph{sklearn} toolkit~\cite{scikit-learn}.
Detailed testbed specifications are listed in Table~\ref{tab:specifications}.

\subsection{Slice Applications}
\label{sec:imple:subsec:slice_app}
We implemented an Android application\footnote{AdaSlicing is fully compatible with other slice applications, as long as the near-RT RIC can periodically retrieve the performance of individual slices.} for each slice.
As AdaSlicing focuses on inter-slice resource allocation, we use only one smartphone as the mobile user for each slice, for the sake of tractability.
It is basically a video streaming application (downlink heavy), where the edge server (collocated in CN desktop) continuously sends video frames to individual mobile users.
Note that, we change the application parameters for each slice (but unknown to AdaSlicing), in terms of the data size of video frames.
The performance metrics of slice applications are 1) throughput and 2) frame-per-second, which are reported to near-RT RIC every second.

\renewcommand{\arraystretch}{1.3}
\begin{table}[!t]
\centering
\caption{The detailed testbed specifications.}
\label{tab:specifications}
\begin{tabular}{|c|c|c|}
\hline
\textbf{Component}    & \textbf{Hardware} & \textbf{Software} \\ \hline
Core Network & Intel Core i7-10700  Desktop       & Open5GS \\ \hline
Open RAN   &  \makecell[c]{Intel Core i7-14700K Desktop}        & OpenAirInterface  \\ \hline
SDR     &  \makecell[c]{Ettus USRP B210}        & UHD v4.5.0.0  \\ \hline
Near-RT RIC  &  Intel Core i7-14700K   Desktop     & FlexRIC v1.0.0   \\ \hline
UEs          &  OnePlus 9 5G         &  Andriod 11  \\ \hline
\end{tabular}
\end{table}

\section{Performance Evaluation}
In this section, we conduct extensive experiments to evaluate the performance of the AdaSlicing system from different perspectives.
The goal is to answer the following questions: 
1) how does AdaSlicing perform as compared to state-of-the-art network slicing systems?
2) how does AdaSlicing optimize online resource allocation in detail? 
3) how does the soft-isolated RAN virtualization improve the resource utilization of slices as compared to existing hard isolation? 
and 4) to what extent, AdaSlicing can adapt to time-varying network dynamics?

We compare AdaSlicing with the following systems:
\begin{itemize}[leftmargin=*]
    \item \textbf{GBO}: GBO uses a global Bayesian optimization to optimize resource orchestration for all slices, under hard-isolated RAN virtualization. To assure the SLA of slices, its objective is penalized by using the the barrier method if their performance requirements are violated. 
    \item \textbf{Atlas}: Atlas~\cite{liu2022atlas} is a state-of-the-art network slicing system, under hard-isolated RAN virtualization. Atlas focuses on the resource allocation of individual slices, which are optimized by using a Bayesian optimization independently. As it does not considered the resource capacity of infrastructures, we slightly modify it to enforce a simple scaling if the resource capacity is exceeded at runtime. 
    \item \textbf{ExSearch}: Exhaustive search (ExSearch) uses the exhaustive search method to optimize the network slicing, under hard-isolated RAN virtualization, where it selects the action with the minimal cost while satisfying the performance requirement of all slices. Note that, it requires the whole action-to-performance dataset to be available in advance, which is impractical in real-world networks.
\end{itemize}

During the performance evaluation, we use the following experiment parameters, which are generally selected based on the realistic network capacity of the testbed.
We create three slices and each slice has one smartphone user.
The performance threshold of all slices $Q_i, \forall i \in \mathcal{I}$ are 12 Megabits-per-second (Mbps) throughput and 10 FPS.
Without loss of generality, we use $U_H=1, U_S=1$ for weighting both svRB and SW.
We use different compression qualities of .jpg in Android applications to stream their video frames to the edge server.
The maximum number of svRBs is $H=12$, where the minimum of 1 svRB is assigned to keep the slice alive.
In the GP, we utilize the \textit{Matern} kernel and \emph{gp-hedge} strategy includes expected improvement (EI), probability of improvement (PI), and low confidence bound (LCB).

\begin{figure}[!t]
	\centering
	\includegraphics[width=2.8in]{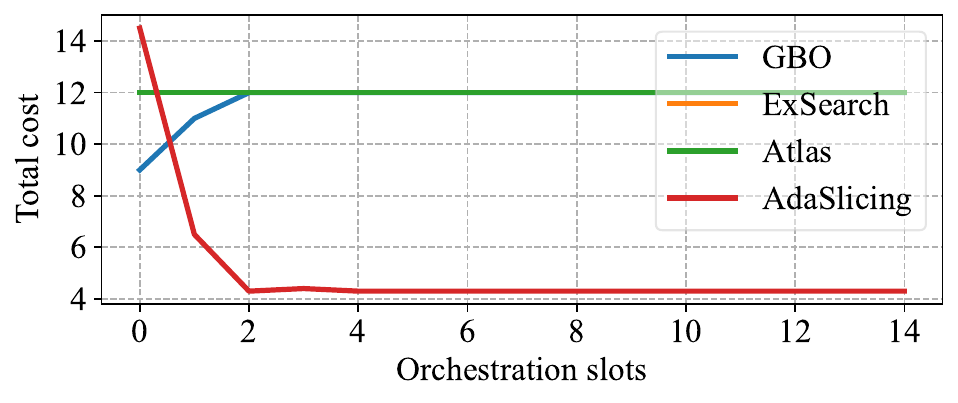}
	\vspace{-0.05in}\caption{\small The convergence of total cost under systems.}
	\label{fig:comparsion_algo}
\end{figure}

\begin{figure}[!t]
	\centering
	\includegraphics[width=2.8in]{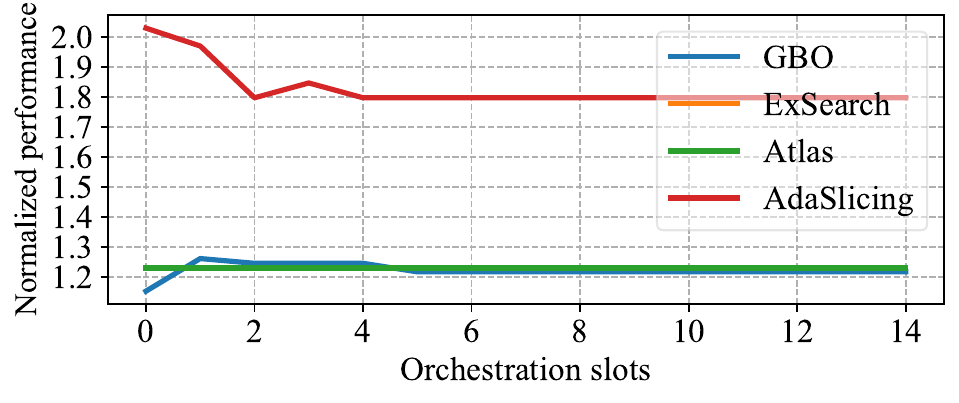}
	\vspace{-0.05in}\caption{\small The convergence of normalized performance under systems.}
	\label{fig:comparsion_Performance_Ratio}
\end{figure}

\subsection{Overall Performance}


\begin{table*}[]
\centering
\caption{Detailed converged orchestration of all systems.}
\label{tb:overall_performance}
\begin{tabular}{c|ccccccc}
\hline
Methods    & Cost        & Individual cost  & svRBs          & SW             & Normalized Performance & Throughput (Mbps)                         & FPS                 \\ \hline
GBO        & 12          & (1, 5, 6)          & (1, 5, 6)        & -              & 1.22                   & (4.40, 12.40, 14.16)          & (16, 18, 12)          \\
ExSearch   & 12          & (4, 4, 4)          & (4, 4, 4)        & -              & 1.23                   & (10.24, 10.24, 10.24)          & (20, 17, 11)          \\
Atlas      & 12          & (4, 4, 4)          & (4, 4, 4)        & -              & 1.23                   & (10.24, 10.24, 10.24)          & (20, 17, 11)          \\
AdaSlicing & \textbf{4.3} & \textbf{(1.1, 2.1, 1.1)} & \textbf{(1, 2, 1)} & \textbf{(0.1, 0.1, 0.1)} & \textbf{1.79}          & \textbf{(17.04, 18.40, 16.80)} & \textbf{(27, 22, 15)} \\ \hline
\end{tabular}
\end{table*}

In this subsection, we show the overall performance (i.e., total cost and slice performance) achieved by all systems.
Fig.~\ref{fig:comparsion_algo} and Fig.~\ref{fig:comparsion_Performance_Ratio} show the convergence of total cost and normalized performance obtained by different systems, respectively.
Here, we define the \emph{normalized performance} of a slice as the ratio between its performance $f(x_i, w_i| s_i)$ and the threshold $Q_i$, where multiple performance metrics will be averaged if applicable. 
We can see that, AdaSlicing achieves a fast convergence speed with only 5 iterations.
This can be attributed to the high sample efficiency of individualized learning agents and the strong robustness of the coordinator in AdaSlicing.
Table~\ref{tb:overall_performance} shows the numerical results of the cost and slice performance after the convergence of all systems.
We can see that, AdaSlicing obtains the lowest cost with the highest normalized performance at the same time, among all systems.
As compared to the state-of-the-art Atlas, AdaSlicing reduces 64.2\% cost, particularly the needed number of svRBs is reduced by 66.7\%, and also improves 45.5\% normalized performance in the meantime.
In other words, although other comparison systems use more virtual resources (i.e., svRBs), their achieved normalized performances are still lower than that of AdaSlicing.
This can be attribute to the soft-isolated RAN virtualization in AdaSlicing, which justifies the necessity of sharing virtual resources among slices at runtime.

\begin{figure}[!t]
	\centering
	\includegraphics[width=2.8in, height=1.7in]{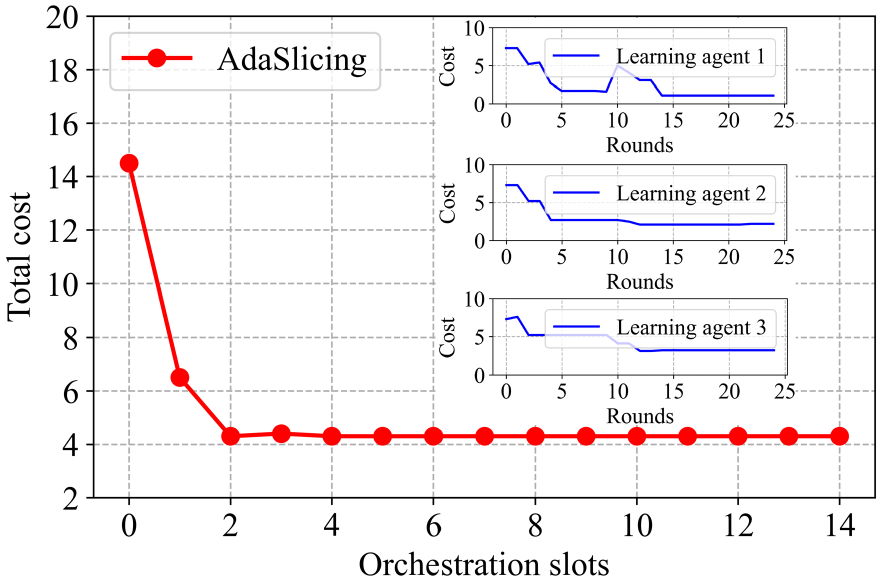}
	\vspace{-0.05in}\caption{\small The convergence of total cost in the AdaSlicing system (the red curve), including three detailed convergence curves under constrained Bayesian optimization (blue curves) at the 4-th orchestration slot.}
	\label{fig:f4_original_cost_history}
\end{figure}

\subsection{AdaSlicing Dissection}
\begin{figure}[!t]
	\centering
	\includegraphics[width=3in]{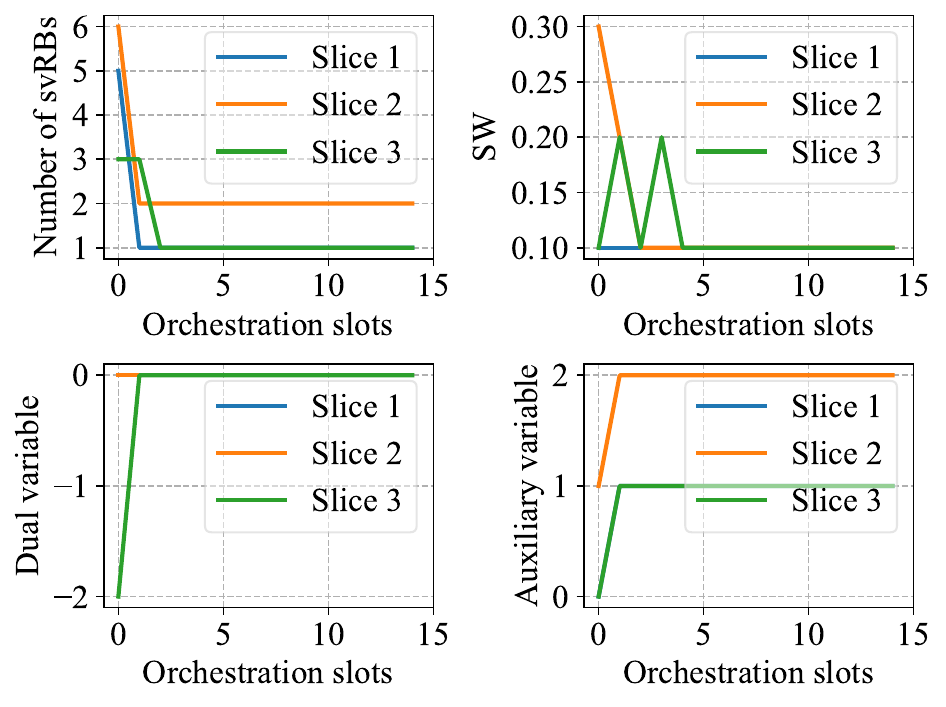}
	\vspace{-0.05in}\caption{\small The convergence of detailed variables.}
	\label{fig:f6_wxyz_foursubplot}
\end{figure}

In this subsection, we dissect the AdaSlicing system to show the details behind its achieved performance in Table~\ref{tb:overall_performance}.
First, Fig.~\ref{fig:f4_original_cost_history} shows the convergence curve of overall cost, and that of all three agents in the $4$th orchestration slot.
It can be seen that, the overall cost starts high and then quickly decreases towards convergence.
This is resulted from both the initialization of auxiliary and dual variables in the coordinator, and the inaccurate representation of learning agents in their early stages.
In AdaSlicing, all learning agents and the coordinator will communicate to achieve the consensus and generate the optimal orchestration action in each orchestration slot.
Moreover, we design to reuse previous experiences in individual learning agents to accelerate their convergence in later orchestration slots, which can be observed in these subplot curves of learning agents.
Fig.~\ref{fig:f6_wxyz_foursubplot} show the detailed convergence of the orchestration actions, and these auxiliary and dual variables.
It can be seen that, after the initialization of dual variables (-5), they are updated to be -2, and then converges to be 0.
This is because that, the optimized number of svRBs by learning agents and the auxiliary variables are exactly equivalent after the second orchestration slot, which leads to no update to dual variables (See Eq.~\ref{update_y}).
These convergence curves verify the effectiveness of the coordinator in coordinating these learning agents in AdaSlicing.

\begin{figure}[!t]
	\centering
	\includegraphics[width=2.6in]{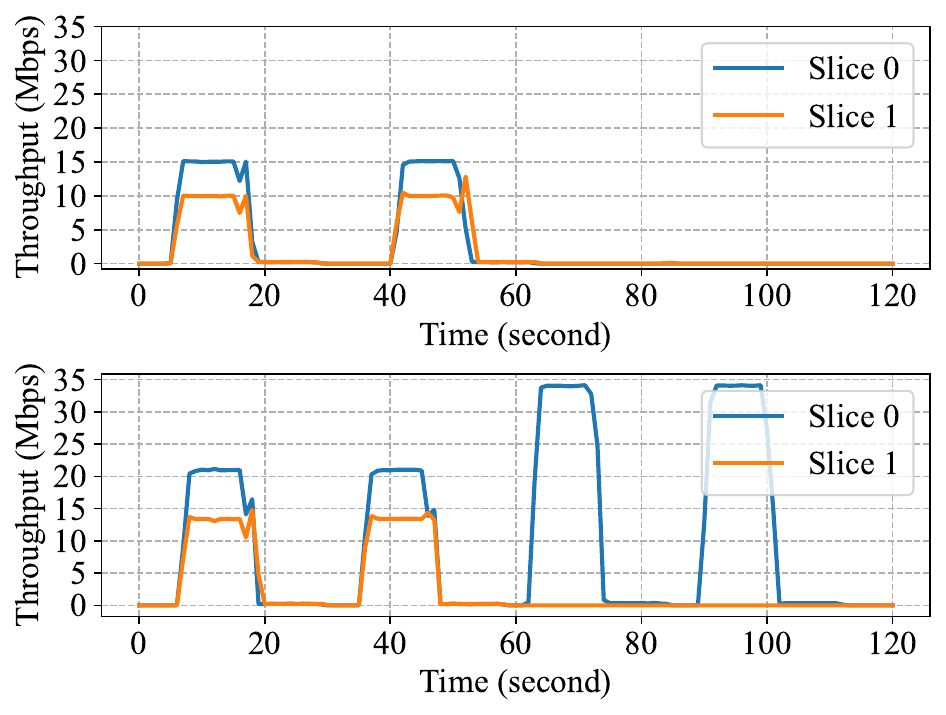}
	\vspace{-0.05in}\caption{\small The experienced downlink throughput under the hard-isolated (top) vs soft-isolated (bottom) RAN virtualization.}
	\label{fig:hardvssoft}
\end{figure}

\subsection{Soft-Isolated RAN Virtualization}
\begin{figure}[!t]
	\centering
	\includegraphics[width=2.6in]{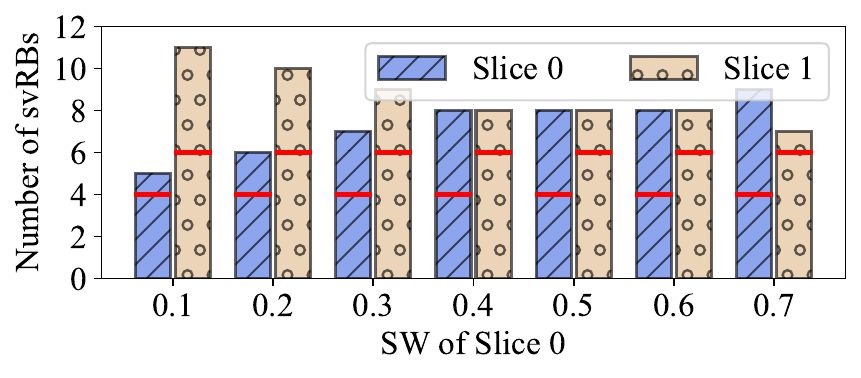}
	\vspace{-0.05in}\caption{\small The impact of SW in sharing virtual resources. Here, the read line in each bar is the originally assigned number of svRBs.}
	\label{fig:changingweight}
\end{figure}

In this subsection, we evaluate the soft-isolated RAN virtualization in AdaSlicing, as compared to existing hard-isolated solutions.
Here, we create two test slices, each has one mobile user, where we measure the throughput by simultaneously starting the online speedtest tool for both slice users.
Fig.~\ref{fig:hardvssoft} shows the experienced throughput of two slices under both soft-isolated and hard-isolated RAN virtualization.
In both experiments, slice 0 and 1 are assigned with [7 RBGs/svRBs, 0.2 SW] and [5 RBGs/svRBs, 0.1 SW], respectively.
It can be seen that, under hard-isolated RAN virtualization (i.e., the top figure), the experienced throughput of both slices are static and generally proportional to their orchestrated number of RBGs.
In contrast, under soft-isolated RAN virtualization (i.e., the button figure), the experienced throughput of both slices are increased, which attributes to unused virtual resources in the RAN.
Besides, the additionally gained throughput of slice 0 is nearly twice than that of slice 1, which proportionally corresponds to their SWs. 
At the later part of the bottom figure (i.e., 62 seconds), we only start the speedtest in the slice 0.
We can see that, its experienced throughput soars up to 33.6 Mbps, which means it enjoys basically all the virtual resources from slice 1 and any other unused resources in RAN.
Fig.~\ref{fig:changingweight} further shows the impact of different SWs in sharing the virtual resources among slices.
Here, we fix the SW of the slice 1 as 0.3, and assign slice 0 and 1 with 4 and 6 svRBs, respectively.
Given the total 16 svRBs in 10MHz RAN, there are 6 svRBs remained unused, which will be shared by slice 0 and 1.
We can see that, the higher SW in the slice 0, its experienced number of svRBs increases accordingly.
For example, when the SW of slice 0 is 0.2, it shares 40\% from the unused 6 svRBs, i.e., 2.4 svRBs, which is grounded to be 2 svRBs.
These results show that, the soft-isolated RAN virtualization can assure the isolation among virtual resources, while improving the resource utilization at runtime.

\begin{figure}[!t]
	\centering
	\includegraphics[width=2.6in]{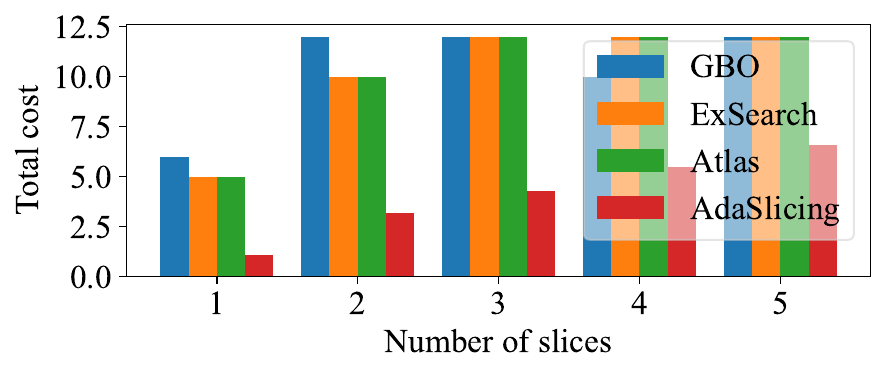}
	\vspace{-0.05in}\caption{\small The total cost under different number of slices.}
	\label{fig:f_final_original_cost_multiple_slices_scaliblity_bar}
\end{figure}
 
\begin{figure}[!t]
	\centering
	\includegraphics[width=2.6in]{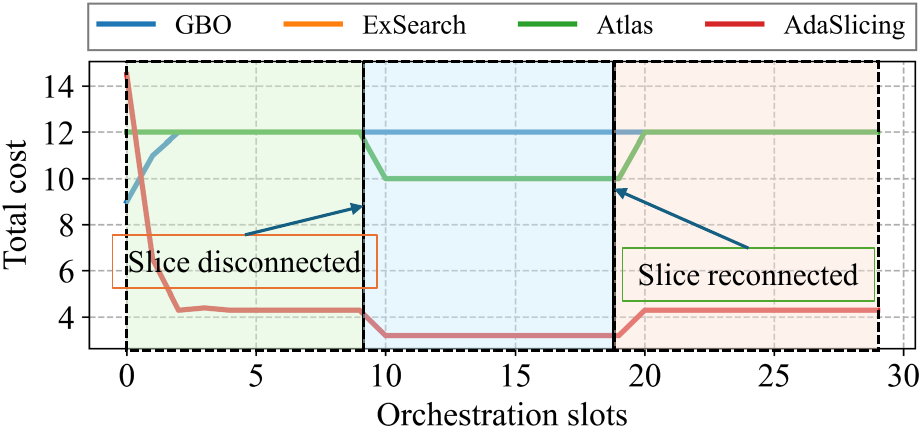}
	\vspace{-0.05in}\caption{\small The total cost under time-varying network dynamics.}
	\label{fig:f_original_cost_history_drop}
\end{figure}

\begin{figure}[!t]
	\centering
	\includegraphics[width=2.6in]{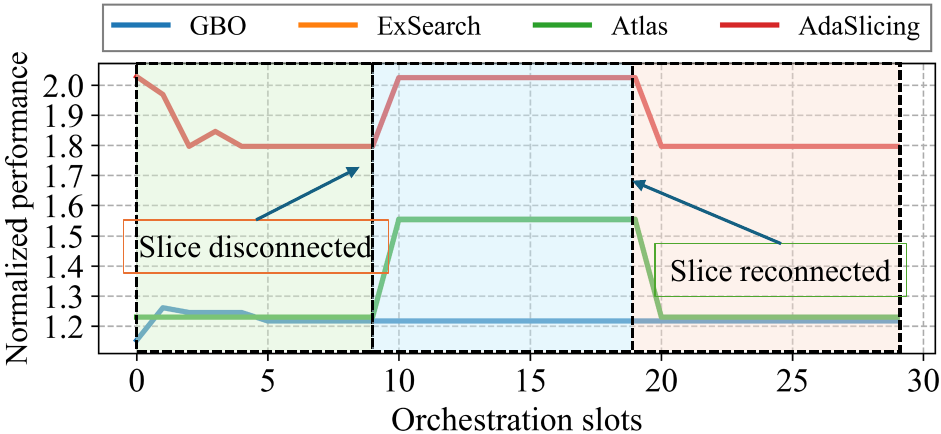}
	\vspace{-0.05in}\caption{\small The normalized performance under time-varying dynamics.}
	\label{fig:f_original_perf_history_drop}
\end{figure}

\subsection{Scalability and Adaptability}
\begin{figure}[!t]
	\centering
	\includegraphics[width=2.6in]{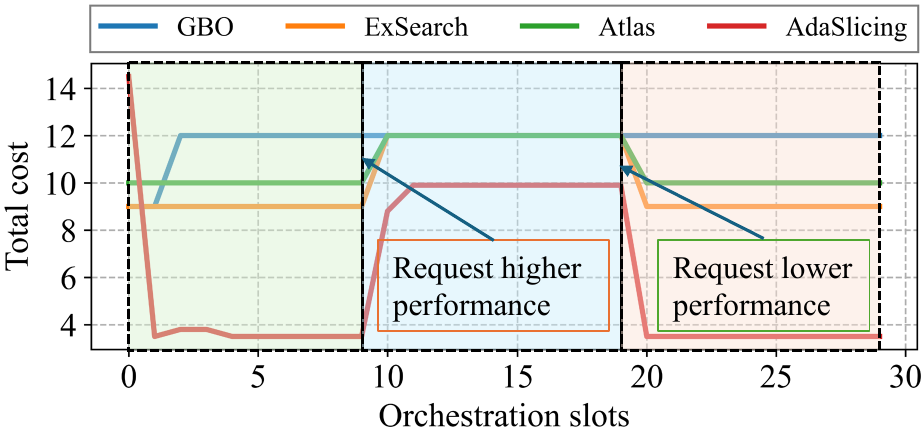}
	\vspace{-0.05in}\caption{\small The total cost under changing slice demands.}
	\label{fig:f_original_cost_history_perf_change}
\end{figure}

\begin{figure}[!t]
	\centering
	\includegraphics[width=2.6in]{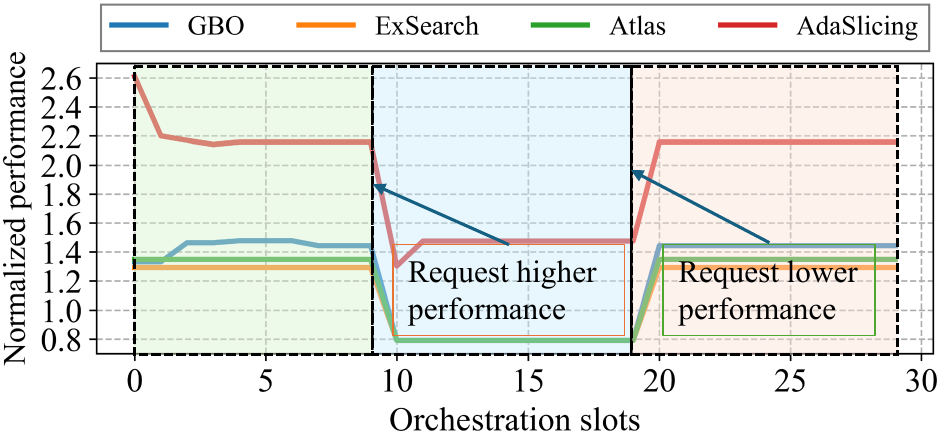}
	\vspace{-0.05in}\caption{\small The normalized performance under changing slice demands.}
	\label{fig:f_original_cost_history_perf_change_ratio}
\end{figure}

In this subsection, we evaluate the AdaSlicing system under different scenarios, in terms of scalability and adaptability.
Fig.~\ref{fig:f_final_original_cost_multiple_slices_scaliblity_bar} shows the total cost of all systems under different number of slices (there are a maximum 5 slices due to the limited capacity of the testbed).
As the number of slices increases in RAN, the more virtual resources are needed to support their users under the slice SLAs.
We can see that, AdaSlicing can achieve the lowest total cost under all scenarios, which verifies the high scalability of AdaSlicing in handling large scale network slicing scenarios.
Here, the total cost of other systems reach to the maximum svRBs in the system, while their normalized performances are decreased up to 53.9\%, if supporting more than 3 slices.
Fig.~\ref{fig:f_original_cost_history_drop} and Fig.~\ref{fig:f_original_perf_history_drop} show the convergence of total cost and normalized performance under time-varying number of active slices. 
Here, we disconnect slice 3 at the 10th orchestration slots and re-connect it back at the 20th orchestration slots, to emulate the network dynamics.
It can be seen that, AdaSlicing quickly adapts to the departure of slice 3 within only one orchestration slot, where the total cost is reduced by 25.6 \% while improving 12.8 \% normalized performance.
This is achieved by adaptively coordinating these auxiliary and dual variables in the coordinator in AdaSlicing.
As slice 3 is reconnected, AdaSlicing can also adapt and converge back to the optima in a few orchestration slots.
In contrast, GBO is designed to accommodate the peak traffic of slices, its total cost cannot be adapted to time-varying network dynamics.
In addition, Fig.~\ref{fig:f_original_cost_history_perf_change} and Fig.~\ref{fig:f_original_cost_history_perf_change_ratio} show the total cost and normalized performance of all systems under time-varying performance threshold of slices.
Here, we set the threshold of slices as [8Mbps, 10FPS], and increase them to [20Mbps, 15FPS] at the 10th orchestration slot, and reset it back at the 20th orchestration slot.
We can see that, AdaSlicing can quickly adapt to this change in a few orchestration slots, by orchestrating more virtual resources to slices.
Here, the decrease of normalized performance in Fig.~\ref{fig:f_original_cost_history_perf_change_ratio} is because we divide the slice performance by the increased threshold.
These experimental results justify the high adaptability of AdaSlicing, in terms of reacting to time-varying network dynamics in real-world networks.

\section{Related Work}

\textbf{Open Radio Access Network.}
Open RAN has shown the increasing momentum in revolutionizing and defining the next generation mobile network~\cite{kak2023aweran}.
FlexRIC~\cite{schmidt2021flexric} is an open-source flexible and efficient software development kit (SDK), and has been gradually adopted to build specialized and multi-service SD-RAN controllers.
HexRIC~\cite{pham2023hexric} is a purpose-built next-generation network controller for the O-RAN ecosystem, featuring with the robust messaging infrastructure and AI/ML operation framework under the architecture of separated control and user plane.
ColO-RAN~\cite{polese2022colo} is the first publicly-available large-scale O-RAN testing framework, that is developed based on the Colosseum wireless network emulator with scaled software-defined radio and computational capabilities.
Open RAN initiatives create unlimited possibilities in next-generation mobile network (e.g., rApps, xApps, dApps), where advanced AI/ML techniques are necessitated to handle ever-increasing complex fine-grained network management.

\textbf{Machine Learning for Networking.}
AI/ML techniques have revealed convincing potential in dealing with complex and time-correlated network systems.
To assure the SLA of end-to-end slices, Liu \emph{et al.} proposed EdgeSlice~\cite{liu2020edgeslice}, a decentralized deep reinforcement learning (DRL) approach, to dynamically orchestrate multi-domain radio and computing resources.
To support increasing computing and storage demand of O-RAN compliant networks, Maxenti \emph{et al.} proposed ScalO-RAN~\cite{maxenti2023scalo}, a control framework to allocate and scale O-RAN applications under the given application-specific latency requirement.
To facilitate spectrum sharing among network operators in O-RAN compliant networks, Bonati \emph{et al.} proposed NeutRAN~\cite{bonati2023neutran}, a zero-touch framework to automate operator onboarding, supported by a new optimization engine and fully virtualized infrastructure.
OrchestRAN~\cite{d2023orchestran} is a network intelligence orchestration framework towards the Open RAN paradigm, and aims to automatically optimize the set of data-driven algorithms while assuring time requirements and avoiding conflicts.
Most policies of existing AI/ML-assisted network managements are trained with offline environments (e.g., simulators) or limited online data observation from real-world networks.
Recent observations revealed that offline policies could suffer from simulation-to-reality discrepancy, leading to non-trivial performance degradation when applied to real-world networks.

\textbf{Online Network Management.}
To address the simulation-to-reality gap, Zhang \emph{et al.}~\cite{zhang2020onrl} proposed OnRL to online update the DRL policy via interacting with real-world networks, to improve the performance of real-time mobile video telephony.
To enable online network configuration in wireless mesh networks (WMNs), Shi \emph{et al.}~\cite{shi2021adapting} proposed a new transfer learning-based algorithm that bridges the simulation-to-reality gap, according to both offline and online datasets.
Hu \emph{et al.}~\cite{hu2023fast} proposed a new neural-assisted algorithm to optimize radio resources to slices, by introducing a DNN to approximate the complex performance function of heterogeneous slices.
Liu \emph{et al.}~\cite{liu2022atlas} proposed Atlas, an online network slicing system, to automate the service configuration of slices with assured slice SLAs, via safe and sample-efficient learn-to-configure approaches.
However, these online learning works heavily rely on parameterized DNN agents with fixed input and output spaces, and cannot efficiently adapt to potential time-varying network dynamics in real-world networks.

\section{Conclusion}
In this paper, we presented AdaSlicing, a new adaptive network slicing system, that online learns to orchestrate virtual resources while efficiently adapting to time-varying network dynamics.
We designed the soft-isolated RAN virtualization that significantly improves the virtual resource utilization of slices without breaking the promise of resource isolation.
We designed the AdaOrch algorithm that minimizes the total operating cost of supporting all slices with assured slice SLAs, via online resource orchestration.
We found that the integration of AI/ML techniques and optimization methods could combine their individual advantages (e.g., high approximation capability and robust convergence) during online resource orchestration in real-world networks.

\section*{Acknowledgement}
This work is supported by the US National Science Foundation under Grant No. 2321699, No. 2333164, and No. 2428427.

\bibliographystyle{IEEEtran}
\bibliography{ref/reference, ref/qiang}

\end{document}